\documentclass[12pt, letterpaper]{article}
\usepackage[utf8]{inputenc}
\usepackage[english]{babel}
\usepackage{wrapfig}
\usepackage{bm}
\usepackage{amsmath}
\newcommand{\STM}{{\mathchoice{}{}{\scriptscriptstyle}{} STM}}
\newcommand{\PC}{{\mathchoice{}{}{\scriptscriptstyle}{} PC}}
\usepackage[pdftex]{graphicx}
\DeclareGraphicsExtensions{.pdf,.jpeg,.png}
\usepackage{authblk}
\usepackage{cite}

\title{Reservoir computing using a spin wave delay line active ring resonator}
\author{Stuart Watt}
\author{Mikhail Kostylev}
\affil{School of Physics and Astrophysics, University of Western Australia, Crawley, W.A. 6009, Australia}

\begin{document}
\maketitle{}

\begin{abstract}
We demonstrate the use of a propagating spin waves for implementing a reservoir computing architecture. Our concept utilises an active ring resonator comprising a magnetic thin film delay line with an integrated feedback loop. These systems exhibit strong nonlinearity and delayed response, two important properties required for an effective reservoir computing implementation. In a simple design, we exploit the electric control of feedback gain to inject input data into the active ring resonator and use a microwave diode to read out the amplitude of the spin waves circulating in the ring. We employ two baseline tasks, namely the short term memory and parity check tasks, to evaluate the suitability of this architecture for processing time series data.
\end{abstract}

\section{Introduction}
The reservoir computer (RC) is a computational framework based on the recursive neural network \cite{jaeger_harnessing_2004,verstraeten_experimental_2007} and as such is suitable for modelling complex dynamical systems. In a RC architecture, a time-dependant data stream is fed into a dynamical system called the ‘reservoir’, whose internal state is governed by nonlinear activation functions and recursive feedback loops. The function of the ‘reservoir’ is to map the input data to a higher dimensional state space in which complex patterns present in the input become linearly-separable and easily determinable using simpler models. Unlike traditional neural network frameworks, training a RC model does not involve modifying the internal properties of the reservoir, and only the readout layer is trained using simple linear regression techniques. Due to this relative ease of training and its remarkable computational power, RC has received great interest in the past decade.\par

In order to be suitable as a RC architecture, the dynamical system must possess two important properties. The first is a nonlinear mapping of the input data to a higher dimensional state space. The second is a ‘fading’ memory where the system state depends not only on the current input but also on past inputs, with the effect of these past inputs ‘fading’ in time. Since RC architectures require only simple training of the read-out layer, they are not restricted to software implementations, and in recent years it has been demonstrated that a wide variety of physical dynamical systems can be exploited for RC implementation with the goal of increasing computation speed and reducing power usage \cite{tanaka_recent_2019, appeltant_information_2011, paquot_optoelectronic_2012, brunner_tutorial:_2018, fujii_harnessing_2017, dion_reservoir_2018, nakajima_exploiting_2014, torrejon_neuromorphic_2017, arai_neural-network_2018, nakane_reservoir_2018}.\par

Spintronic based architectures \cite{torrejon_neuromorphic_2017, arai_neural-network_2018, nakane_reservoir_2018} are promising candidates for practical RC applications due to their low power usage, strong nonlinearity arising from magnetisation dynamics, and their ability to be scaled down to small sizes. In particular, many studies have already been conducted on spin-torque nano-oscillators with experimental results \cite{torrejon_neuromorphic_2017, arai_neural-network_2018, romera_vowel_2018, tsunegi_evaluation_2018, tsunegi_physical_2019, riou_temporal_2019} demonstrating promising performance as RC implementations. Further simulations \cite{furuta_macromagnetic_2018, kanao_reservoir_2019} indicate the potential for improved performance by using arrays of coupled oscillators. Other proposed spintronic implementations are systems utilising magnetic skyrmion memristors \cite{jiang_physical_2019} and spin-wave (SW) interference in garnet films \cite{nakane_reservoir_2018}. In the present work we propose an alternative spintronic RC architecture also based on a SW propagation. The physical set-up, shown in Fig 1(a), is a magnetic-film ring resonator consisting of a spin-wave delay line with an amplified feedback loop. This system utilises the delay and nonlinear behaviour of travelling spin waves in magnetic films and has been used in many studies to explore nonlinear behaviour such as soliton formation, modulational instability and chaotic behaviour \cite{wu_nonlinear_2010}. First we describe the SW active ring resonator system and demonstrate how it can be used for RC implementation. We then briefly cover the principles of RC followed by experimental evaluation of the system’s performance as a RC using benchmark tasks. 

\section{Spin-wave delay line active ring resonator}

The delay line is made up of a 4 mm wide and 22 mm long yttrium-iron garnet (YIG) film with thickness of 43 $\mu$m. The film sits on two short-circuited, 0.5 mm microstrip antennae separated by a distance of 12 mm. The feedback loop consists of a broadband microwave amplifier, an adjustable attenuator, a directional coupler and a PIN-diode microwave switch (MS). A magnetic field is applied parallel to the antennae and perpendicular to the spin wave propagation direction which allows the excitation of magnetostatic surface waves (MSSW) which propagate between the antennae. The dispersion relation for these waves is given by \cite{stancil_spin_2009}

\begin{figure*}[!t]
	\centering
	\includegraphics[width=1\textwidth]{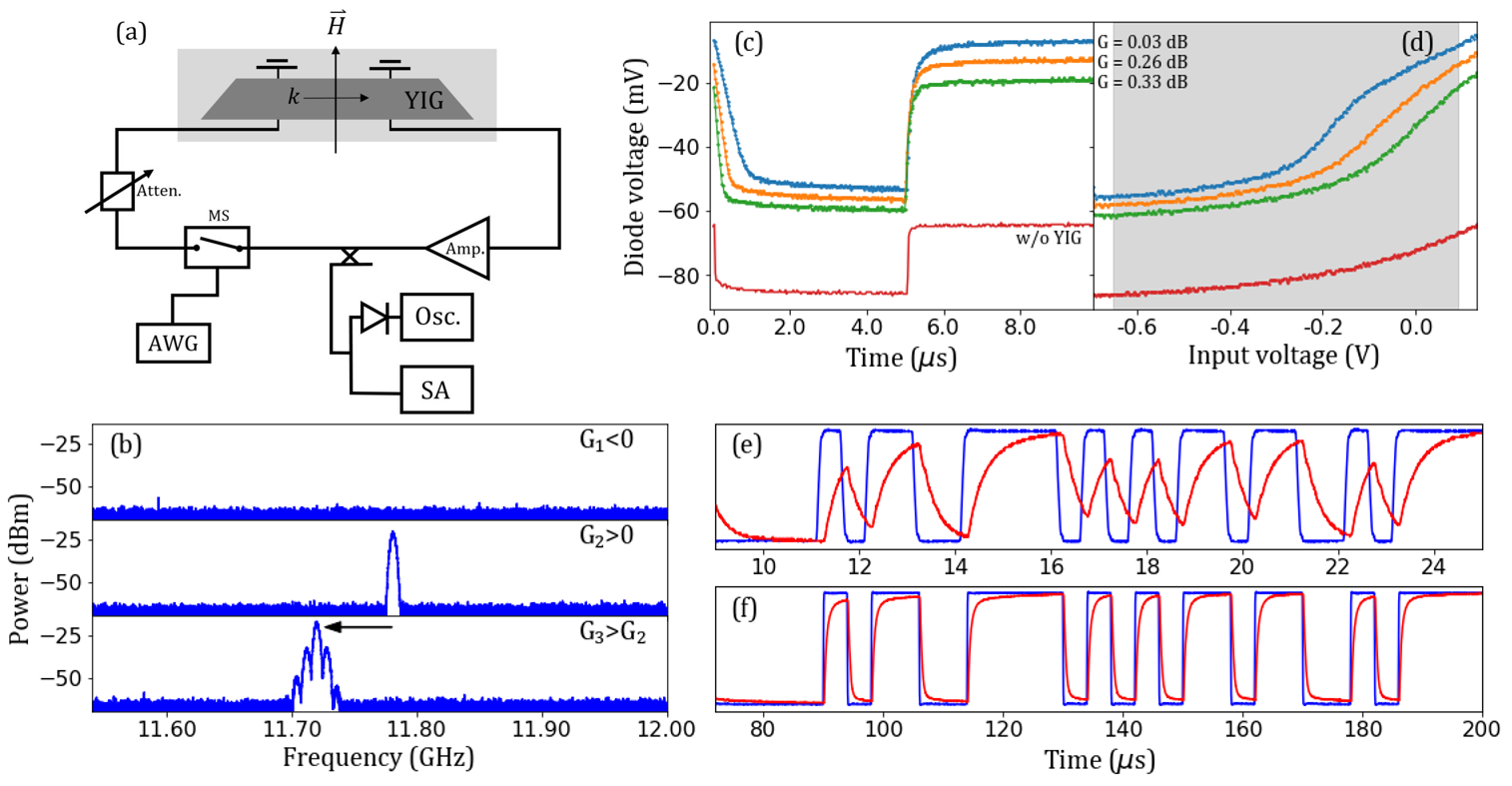}
	\hfil
	\caption{(a) Schematic diagram of the spin wave active ring active ring resonator system. Components are described in the text. (b) Frequency spectrum of the active ring for various gains. (c) Resonator state response to sudden change (step function) in the microwave switch input bias voltage for various gains. (d) Resonator state response to microwave switch input bias voltage for various gains. The grey region indicates the computing range of input bias voltage. (e) The system input (blue) and the diode voltage output (red) in response to a binary input sequence with an input time interval of 0.5 $\mu$s. (f) Same as for (e) but with an input time interval of 4 $\mu$s.}
	\label{fig:FIG1}
\end{figure*}

\begin{equation}
\omega(H, k) = \gamma\sqrt{H(H+4 \pi M_{s})+\frac{(4 \pi M_{s})^{2}}{4}(1-e^{-2 k L})}.
\end{equation}

In this equation $H$ is the magnetic field and $k$ is the SW wavenumber. The constants $\gamma$, $4 \pi M_{s}$ and $L$ are the gyromagnetic ratio, the saturation magnetisation and the film thickness, which have values 2.8 MHz/Oe, 1750 Oe and 43 $\mu$m respectively.  The travelling spin waves are excited at the first antenna by the microwave current-induced dipole field and detected at the other antennae through inductive coupling.\par

The signal from the output antenna is then amplified by some gain and fed back into the input. The SW delay line introduces a time delay of the signal passing from the input to the output antennae resulting in a phase shift of the microwave signal across the YIG strip. The gain, $G$, in the feedback loop is controlled using the adjustable attenuator. If the gain is sufficient to compensate the losses occurring in the YIG film, then a signal can begin to circulate in the ring. Assuming the phase shift resulting from the electrical components is negligible, resonant frequency eigenmodes will occur in the ring when the SW wavenumber satisfies the condition, $k_{res}d=2 \pi n$, where $d$ is the antennae separation. The eigenmodes which are excited depend on the frequency characteristic of the delay line. \par

Shown graphically in Fig 1(b) is the process of ring eigenmode excitation measured using a spectrum analyser (SA). At low gain, delay line losses dominate and the spin-waves do not have enough power to propagate between the two antennae. Above a certain threshold (defined as $G$=0) where the gain is sufficient to compensate the losses, thermally excited magnons with wavenumbers corresponding to the eigenmode with the lowest loss will be resonantly amplified in the ring (G2 in Fig 1(b)). As the gain increases the next eigenmode with the lowest loss is excited, and so on (G3). The precession angle increases with SW power, effectively reducing $4 \pi M_{s}$ and shifting the dispersion curve downward. This results in a shift of the eigenmode frequency.\par

These systems exhibit a host of nonlinear behaviour which we propose can be exploited for physical RC. In particular, in this work we exploit the nonlinearity of SW damping at high microwave power \cite{scott_nonlinear_2004}, for which the active ring resonator can be operated in the single mode regime (G2 in Fig 1(b)). The system state is read by measuring the amplitude of the circulating signal using a microwave tunnel diode with a fast oscilloscope.\par

The parameters which control the dynamics of this system are the ring gain and the magnetic field strength. In this work we employ a static magnetic field and use a variable ring gain as the method of data input. This is done by using the microwave switch which varies the ring gain by attenuating the microwave signal according to the DC bias voltage supplied by an arbitrary waveform generator (AWG). Fig 1(d) shows the ring state in response to the microwave switch input voltage at various levels of the bias ring gain $G$. The grey shaded region shows the input voltage range used in this work. The resonator state depends nonlinearly on the input voltage, satisfying the first condition required for RC implementation. Furthermore, the nonlinearity depends on the value of $G$, with the more nonlinear behaviour occurring closer to the auto-oscillation onset threshold ($G$=0). It should be noted, that the nonlinearity has contributions from both the YIG film and also the switch/diode components. The bottom curve (labelled ‘w/o YIG’ in Fig 1(d)) shows the response of the switch/diode components acting alone on a constant microwave signal, showing that they too result in a nonlinear diode voltage.  \par

The second property required by a RC implementation is the ‘fading’ memory property. In this system, the ‘fading’ memory is achieved through the delayed response of the system to a given input. The resonator state response to a step function of the input voltage is shown in Fig 1(c). The time for the resonator to reach a steady state, where the circulating ring signal has a constant amplitude, also depends on the value of $G$, with longer relaxation times occurring at gains closer to the threshold. The round trip time of the signal in the loop is approximately 50 ns, and the signal must travel around the loop multiple times before reaching a steady state. Thus control of the gain in the ring allows partial tuning of both the nonlinearity and response time.\par

\section{Reservoir computing performance}

The traditional RC model has three main components - the input, the reservoir and the read-out\cite{montavon_practical_2012}. The system state at discrete time step $T$ is a vector containing the values of each node in the reservoir determined by the update equation

\begin{equation}
\vec{x}(t)=f( \bm{W^{in}} \vec{u}(t) + \bm{W} \vec{x}(t-1)).
\end{equation}

The matrices $\bm{W^{in}}$ and $\bm{W}$ determine the weights for the input layer and internal connections respectively. $\vec{u}(t)$ is the vector of input values, and $f$ is the  nonlinear function which governs the response of the system. $f$ and $\bm{W}$ are determined by the physical parameters of the RC implementation, and the output of the system is obtained as a weighted sum of the reservoir states

\begin{equation}
\vec{y}_{out}(t)=\bm{W^{out}} \vec{x}(t).
\end{equation}

During the training phase, the weights in $\bm{W^{out}}$ are adjusted to reduce the mean squared error between the RC output and the target output. Since in RC, only the output weight matrix $\bm{W^{out}}$ is trained, all of the input data can be passed through the reservoir at once, and then the training and evaluation are performed off-line later. Defining $\bf{Y}$ and $\bf{X}$ as matrices containing the target and reservoir states for all time steps, the optimal $\bm{W^{out}}$ is obtained by taking the product $\bf{Y} \bf{X}^{-1}$.

In order to process continuous time signals, one must first sample the input, and then each discrete value is exposed to the system for a time interval of $\theta^{int}$.This time interval can be chosen to control the level of ‘fading’ memory in the system. Examples of the active ring resonator state output to the input binary sequence are shown in Fig 1(e) and (f) for input intervals of 0.5 $\mu$s and 4 $\mu$s respectively. For larger values of $\theta^{int}$ the system has sufficient time to reach a steady state which depends less on the past inputs, reducing the ‘fading’ memory. Unlike traditional RC, where the reservoir is comprised of many nodes, the proposed system is effectively a single node with a single input and single recursive feedback loop. In order to increase the dimensionality of the reservoir, the resonator output is sampled into N ‘virtual’ nodes\cite{appeltant_information_2011} for each input time interval with separation $\theta^{int}/N$, creating the reservoir state vector $\vec{x}(t)$. As is typical in RC applications, the input value for a given time step and an additional bias term (equal to -0.2 for all time steps) are also added to the state vector and used for training and prediction. Thus there is a total of $N+2$ values used to compute the system output for each time step. \par

We employ two popular metrics to evaluate the performance of the active ring resonator system as a RC. These are the short-term memory (STM)\cite{jaeger_tutorial_2002} and parity check (PC)\cite{furuta_macromagnetic_2018, bertschinger_real-time_2004} tasks. For both, the system state at a given time T is measured in response to a random binary input $u(T) \in [0,1]$. A sequence of 2100 time steps is input to the system where the first 100 steps are used to ‘washout’ any transient behaviour in the system which may depend on the arbitrary initial state. The next 1000 time steps are used for training $\bm{W^{out}}$  and the final 1000 steps for evaluation. In order to input the sequence to the active ring resonator, the inputs [0,1] are converted to corresponding input bias voltages [-650 mV, 90 mV], which correspond to the boundaries of the input range shown in Fig 1(d). For each step in the input sequence, the resonator output is split into N=20 nodes.

The STM task provides a measure of the system’s ability to predict previous inputs given the current output. In this task, the target output for each time step is simply the input at some previous time in the past, $y_{\STM}(T, \tau) = u(T - \tau)$, where $\tau$ is the time delay. Following the standard training procedure, the output weight matrix $\bm{W^{out}}$ is determined using the training set. The system then makes predictions on the test set, and the square of the correlation coefficient between the ideal targets and the model predictions is determined as \cite{furuta_macromagnetic_2018}

\begin{figure}[!t]
	\centering
	\includegraphics[width=3in]{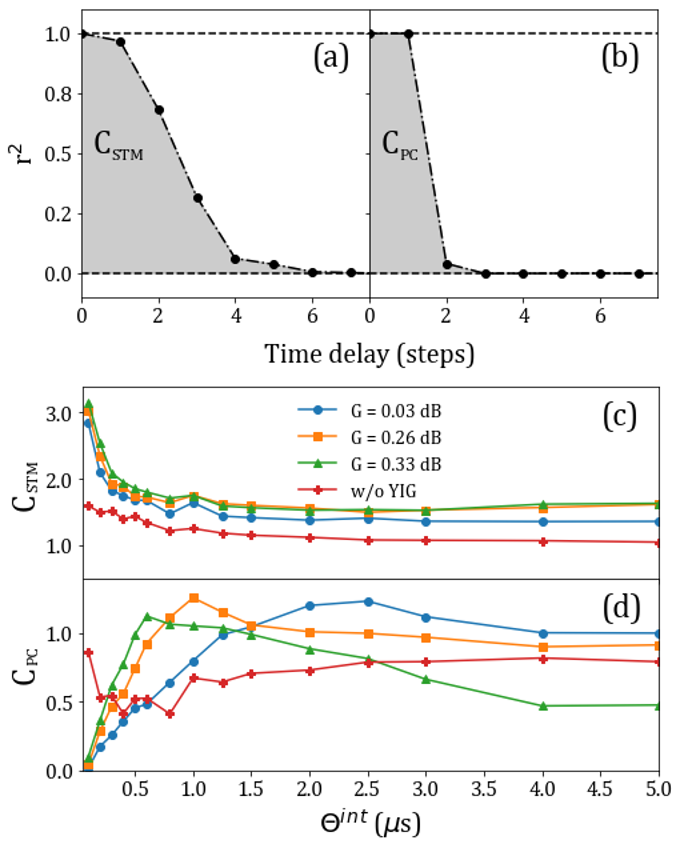}
	\hfil
	\caption{(a-b) Correlation coefficient squared for increasing delay (forgetting curves) for the STM and PC tasks. Data here are for $\theta^{int}$ equal to 0.3 and 1 $\mu$s respectively. (c) The STM capacity versus $\theta^{int}$ for different gain levels. The data ‘w/o YIG’ corresponds to the STM capacity of the electronic components alone. (d) Same as (c) but for the PC task}
	\label{fig:FIG2}
\end{figure}

\begin{equation}
r_{\STM}(\tau)^{2} = \frac{Cov(y_{out}(T), y_{\STM}(T, \tau))^{2}}{Cov(y_{out}(T)) \times Cov(y_{\STM}(T, \tau))}.
\end{equation}

Here $Cov(A,B)$ is the covariance between some vectors A and B, and $Cov(A) \cong Cov(A, A)$. $r_{\STM}(tau)^{2}$ takes values between 0 and 1, where the value of 1 indicates perfect replication of the targets. The STM capacity is then calculated by taking the sum of $r_{\STM}(tau)^{2}$ over the range of delays

\begin{equation}
C = \sum_{\tau=1}^{\tau_{max}=20} r_{\STM}(\tau)^{2}.
\label{capacity}
\end{equation}

While the STM task is a good characterisation a RC’s capability to store memory, it does not require a nonlinear system in order to yield high STM capacities and thus is not a good metric to evaluate the nonlinearity performance of the RC. The PC task is a non-linearly separable task which requires both ‘fading’ memory and nonlinearity. The target outputs used in this task are determined by taking the parity of the sum of the consecutive binary inputs up to some time delay in the past

\begin{equation}
y_{\PC}(T, \tau) = PARITY(u(T-\tau), u(T-\tau+1),...,u(T)).
\end{equation}

Here the PARITY operation returns the parity (0 for even, 1 for odd) of the sum of the parameters in the brackets, and as such the target itself is also a sequence of binary values. It should be noted that the evaluation of the nonlinear characteristics is not restricted to the PC task; one may also choose a task based on some other arbitrary binary operation acting on past inputs. Evaluating the performance on the PC task is carried out in the same manner as for the STM task, with the PC capacity similarly determined using Eq. \ref{capacity}. Since the STM and PC tasks both use binary sequences, only one input sequence is needed to compute both $C_{\STM}$ and $C_{\PC}$. \par

The results of our experiments are shown in Fig. 2. From the figure, one sees that as the delay time increases, the ability of the system to correctly predict the targets reduces. This can best be described using ‘forgetting’ curves \cite{jaeger_tutorial_2002} as shown in Fig 2 (a-b), which plot the square of the correlation coefficient against the delay time. The respective capacity is then calculated as the area under the curves. The plot of $C_{\STM}$ versus the input time interval is shown in Fig 2(c). One sees that the active ring system performs better than using electronic components alone for all $\theta^{int}$, indicating the importance of the feedback loop for realizing the ‘fading’ memory. The $C_{\STM}$ increases for smaller input time intervals, which intuitively can be explained since the system cannot reach a steady state and the final output depends on the previous system state. The minor improvement in $C_{\STM}$ with increasing gain is likely the result of reduced noise as the ring resonator moves away from the auto-oscillation threshold ($G$=0). At the lowest input time interval, the $C_{\STM}$ reaches its maximum value of 3.16. \par

The performance of the active ring evaluated using the PC task (fig 2(d)) shows significant dependence on gain, with $C_{\PC}$ reaching its maximum at larger values of $\theta^{int}$ for lower feedback gains. Similar to the STM task, the active ring performs significantly better than the electronic components alone except at higher gains and time intervals. We obtain a maximum $C_{\PC}$ of 1.27 for an input time of 1 $\mu$s. These STM and PC results are comparable to those from Ref. \cite{tsunegi_physical_2019} for a single spin-torque oscillator using a single binary sequence output. Ref. \cite{tsunegi_physical_2019} shows how the results can be further improved by averaging over multiple repeated sequences. 

\section{Conclusion}

The results we obtained suggest that the system we propose has potential for development in the growing field of physical RC implementation. As well as being a simple design, this concept requires little additional pre- or post-processing of the input or output. Furthermore, YIG is one of the most nonlinear materials available and is extremely robust. The closest competitor of our concept, spin-torque nano-oscillators, require injection of currents with high densities for their operation and can be damaged by excess currents of static voltages. This technical problem is naturally solved for our concept of a dielectric-based device. Furthermore, low-magnetic-loss YIG films are now available in a very large range of thicknesses – from several nanometers to several tens of micrometers. With a decrease in the film thickness, the SW group velocity decreases leading to an increase in the delay time per unit film length. In addition, for thinner films, one may also expect that smaller microwave powers (i.e. smaller amplification gains in the feedback loop) are needed to reach the nonlinearity threshold for spin waves. This shows that our concept can easily be scaled down for integration on chips and one may even expect improvement of performance with scaling down. 

\section{Acknowledgments}
Research Collaboration Award and Vice Chancellor’s Senior Research Award from the University of Western Australia are acknowledged. The work of S. Watt was supported by the Australian Government Research Training Program.

\bibliographystyle{IEEEtran}
\bibliography{Reservoir_Computing}

\end{document}